\newtheorem{theorem}{Theorem}
\newcommand{\norm}[1]{\left\lVert#1\right\rVert}
\newcommand\abs[1]{\left|#1\right|}

\documentclass[journal]{IEEEtran}
\usepackage{enumitem}
\ifCLASSINFOpdf
   \usepackage[pdftex]{graphicx}
\else
  \usepackage[dvips]{graphicx}
\fi
\usepackage[cmex10]{amsmath}
\usepackage{amssymb}

\begin{document}
%
\title{Online Wideband Spectrum Sensing Using Sparsity}
%
%
%
\author{Lampros~Flokas
        and~Petros~Maragos,~\IEEEmembership{Fellow,~IEEE}
\thanks{Lampros Flokas is with the Columbia University in the City of New York, Department of Computer Science, New York, USA, \mbox{Email: lamflokas@cs.columbia.edu}.\par Petros Maragos is with the National Tech. University of Athens, School of ECE, Greece, Email: \mbox{maragos@cs.ntua.gr}}
}
\maketitle

\begin{abstract}
Wideband spectrum sensing is an essential part of cognitive radio systems. Exact spectrum estimation is usually inefficient as it requires sampling rates at or above the Nyquist rate. Using prior information on the structure of the signal could allow near exact reconstruction at much lower sampling rates. Sparsity of the sampled signal in the frequency domain is one of the popular priors studied for cognitive radio applications. Reconstruction of signals under sparsity assumptions has been studied rigorously by researchers in the field of Compressed Sensing (CS). CS algorithms that operate on batches of samples are known to be robust but can be computationally costly, making them unsuitable for cheap low power cognitive radio devices that require spectrum sensing in real time. On the other hand, online algorithms that are based on variations of the Least Mean Squares (LMS) algorithm have very simple updates so they are computationally efficient and can easily adapt in real time to changes of the underlying spectrum. In this paper we will present two variations of the LMS algorithm that enforce sparsity in the estimated spectrum given an upper bound on the number of non-zero coefficients. Assuming that the number of non-zero elements in the spectrum is known we show that under conditions the hard threshold operation can only reduce the error of our estimation. We will also show that we can estimate the number of non-zero elements of the spectrum at each iteration based on our online estimations. Finally, we numerically compare our algorithm with other online sparsity-inducing algorithms in the literature.  
\end{abstract}

\begin{IEEEkeywords}
signal processing, sparse representations, LMS, cognitive radio.
\end{IEEEkeywords}

%
\IEEEpeerreviewmaketitle

\section{Introduction}
%
%
%
%
\IEEEPARstart{W}{ireless} telecommunications spectrum is a limited resource and with the rapid increase of telecommunication applications, static allocation of spectrum for each case is not a viable solution. Additionally, static allocation of spectrum is also not effective as the primary users of the spectrum may use it from time to time and only in some locations. To overcome this limitation cognitive radio devices try to dynamically manage the spectrum by detecting which part of the spectrum is unused by its primary users and temporarily using it for their own needs. 
\par In order to be effective, these devices would need to check a wide band of frequencies to increase the possibility of finding unused spectrum. If cognitive devices used sampling rates that are equal or above the Nyquist rate, their cost would be prohibitive for most applications. In order to reduce the sampling rate needed as well as the computational effort, we will need to use some prior information on the structure of the received signal. This prior is that the same one that enables the usage of cognitive radio devices in the first place: Primary users do not use their share of the spectrum all the time so the received signal should be sparse in the frequency domain.
\par The area of compressed sensing (CS) has provided several celebrated algorithms for the reconstruction of undersampled signals with sparse representations \cite{donoho2003optimally,candes2006robust,candes2006stable}. Classic algorithms of CS assume a batch setting where the device is assumed to collect a number of observations and operate on them  in an iterative manner. Therefore it is of great importance to provide algorithms that reduce the number of iterations needed in order to reduce the computational burden on the cognitive radio devices and provide real time spectrum estimations. CS based approaches have been adapted by many researchers in the area of spectrum sensing for cognitive radio applications \cite{Tian07,Tian08,Zhang2011,Zeinalkhani2012,Sun2013,Zhang2016,Sharma2016,Qin2017}.
\par On the other hand, online algorithms, based on variations of Least Mean Squares introduced by Widrow and Hoff \cite{WidrowAd}, have also been adapted for the CS setting. Algorithms like the ones presented in \cite{martin2002exploiting,Gu2009,Them14,Chen09,angelosante2010online,babadi2010sparls,Yu2015} have been shown to estimate sparse signals with faster convergence and smaller steady state errors than methods that do not exploit sparsity. Additionally, they have much simpler updates based on a single sample at a time. This allows them not only to be more computationally efficient but also to be adaptive to the changes of the estimated signal.   
\par Here we will propose two new variations of the classical LMS algorithm. The first is a variation of the Zero Attracting LMS \cite{Chen09} that does not penalize the $s$ algebraically largest coeffients of the estimation where $s$ is an upper bound on the number of non-zero elements in the estimated vector. The second one alternates the standard LMS update with shrinkage using a hard threshold operator. The hard threshold operator will keep the $s$ algebraicly largest components of the estimated vector where $s$ is again an upper bound on the number of non-zero elements in the estimated vector. This algorithm is the online version of the iterative hard thresholding studied in \cite{Blum09} and \cite{Blum10} and \cite{garg2009gradient}. The sparsity of the estimated vector or even an upper bound on it may not be known in advance so we will also propose a way to estimate it in an adaptive manner. Even though we are going to apply the proposed algorithms for the problem of spectrum estimation, they can also be applied in other telecommunications and general machine learning applications where the incoming signals have a known sparse representation.  
\par The structure of the paper is as follows. In Section 2 we will define the problem of sparse spectrum reconstruction using below Nyquist rate sampling frequencies. In Section 3 we will present the properties of online sparsity aware estimation techniques in the literature and in Section 4 we will introduce our hard thresholding based algorithms. In Section 5 numerical simulations comparing our algorithm with other sparsity aware algorithms are provided. Finally, Section 6 contains concluding remarks and discusses possible directions for future research.
\section{Problem statement}
\par Let $\mathbf{z} \in \mathbb{R}^N$ be the full signal that the cognitive radio device would receive if it was sampling it at the Nyquist rate. We would like to undersample $\mathbf{z}$, taking just $M$ observations from $\mathbf{z}$ where $M<N$. Let us call $\mathbf{U}$ the undersampling matrix whose rows are a subset of the rows of the identity matrix including only the rows where the respective observation of $\mathbf{z}$ is sampled. Let us call $\mathbf{y}=[y_0,y_1,\dots,y_{M-1}]^T  \in \mathbb{R}^M$ the resulting vector. If each observation $y_i$ is corrupted by an additive error term $v_i$ and $\mathbf{v}=[v_0,v_1,\dots,v_{M-1}]^T \in \mathbb{R}^M$ Then we obviously have that 
\begin{equation}
\mathbf{y}=\mathbf{U}\mathbf{z} + \mathbf{v}
\end{equation}
Of course without any further assumptions the lost information cannot be recovered and important information about the spectrum of $\mathbf{z}$ cannot be estimated. However, in our case we can assume that the underlying spectrum of the signal is sparse as a large portion of the spectrum  will be left unused by its primary users. Let  $\mathbf{w} \in \mathbb{C}^N$ be the complex vector representing the Discrete Fourier Transform (DFT) of $\mathbf{z}$ and $\mathbf{\Phi}$ be the Inverse Discrete Fourier Transform (IDFT) matrix so that $\mathbf{z}=\mathbf{\Phi}\mathbf{w}$. Given our assumption on the sparsity of the spectrum of $\mathbf{z}$, we have that $\mathbf{w}$ is a sparse vector and therefore we are interested in solving the following problem:
\begin{equation} \label{pr:ell0}
\min \norm{\mathbf{w}}_0: \norm{\mathbf{y}-(\mathbf{U}\mathbf{\Phi})\mathbf{w}}_2 \leq \delta
\end{equation}
where the $\ell_0$ norm is the count of non-zero elements of the vector i.e $\norm{\mathbf{w}}_0=|\mathrm{support}(\mathbf{w})|$, where $\mathrm{support}(\mathbf{w})=\{ i \in \{0,1,..,N-1\}: w_i \neq 0\}$,  $|S|$ denotes the cardinality of set $S$ and $\delta$ is an upper bound on $\norm{\mathbf{v}}_2$. In general this problem is NP-hard and therefore computationally intractable. However, researchers in the area of CS have developed several algorithms that recover the solution of problem described by \ref{pr:ell0} when the matrix $\mathbf{U}\mathbf{\Phi}$ satisfies the Restricted Isometry Property and vector $\mathbf{w}$ is sparse enough. Out of all the algorithms probably the most popular one is Lasso regression. One of its equivalent formulations is
\begin{equation} \label{pr:ell1}
\min \norm{\mathbf{w}}_1: \norm{\mathbf{y}-(\mathbf{U}\mathbf{\Phi})\mathbf{w}}_2 \leq \delta
\end{equation} 
The resulting optimization problem can be solved with standard convex optimization methods. The limiting factor for Lasso and other classical approaches to CS is that they may require several iterations to converge to the optimal solution. This makes them unsuitable for the low power cognitive radio devices that need real time spectrum estimations in potentially highly volatile settings.
\par In contrast, online estimation algorithms have much simpler update rules that involve one sample at a time and are robust to changes in the estimated signal. In the online setting there is a stream of measurements of $\mathbf{y}$ and the corresponding rows of $\mathbf{\Phi}$ that are fed one by one to the online algorithm. There are at least two options when it comes to forming this stream.
\begin{enumerate}
\item The first option is to use an online algorithm as a drop in replacement of a batch algorithm. We can collect $M$ out of $N$ samples of the signal and feed them one by one to the online algorithm. Of course the online algorithm may not converge in a single pass over the data so we can augment the stream by feeding the same measurements  to the algorithm multiple times in order to achieve convergence. The advantage of the online algorithms over batch algorithms in this setting is that they have simpler update rules than their batch counterparts and so they could be more easily implementable in low power cognitive radio devices.  
\item The second option is to form a stream by continuously incorporating new measurements. One way to do this is to split the incoming signal in non overlapping windows of length $N$, randomly sample $M$ measurements in each window and feed the resulting measurements to the online algorithm. The advantage of the online algorithms over batch algorithms in this setting is that they can track the spectrum changes in the signal in real time.
\end{enumerate}
In Section 5 we shall provide experimental results for both settings.   
\section{Related Work} 
\subsection{The LMS algorithm}
The algorithms proposed in this paper are based on the updates of the LMS algorithm. To better understand the procedure we review the steps of the classical LMS algorithm. Let $y(n)$ be a sequence of observations of the output of a system following the model
\begin{equation} \label{eq:lms}
y(n)=\mathbf{w}^H\mathbf{x}(n)+v(n)
\end{equation}
where $\mathbf{w}=[w_0,w_1,\dots,w_{N-1}]^T \in \mathbb{C}^N$  is the parameter vector to be estimated, $\mathbf{x}(n)\in \mathbb{C}^N$ is taken from the rows of the $\mathbf{\Phi}^*$ that correspond to the observed samples and $v(n)$ is the additive observation noise. Let also $\mathbf{w}(n)$ be the estimation we have up to time $n$ for the unknown vector $\mathbf{w}$ and $e(n)$ be the sample error. Then
\begin{equation} \label{eq:error}
e(n)=y(n)-\mathbf{w}^H(n)\mathbf{x}(n)
\end{equation}
\par The LMS update rule is recursive and produces a new estimation given the previous one, following the rule
\begin{equation}
\mathbf{w}(n+1)=\mathbf{w}(n)+\mu e^*(n)\mathbf{x}(n)
\end{equation}
where $\mu$ is a an appropriately chosen constant. If $\mathbf{R}_x=\mathbb{E}[\mathbf{x}(n)\mathbf{x}^H(n)]$ is the uncentered covariance matrix of $\mathbf{x}(n)$, here assumed constant over time, and $\lambda_{max}$ is its maximum eigenvalue then \cite{widrow1971adaptive} shows that LMS will converge in the mean sense if:
\begin{equation} \label{eq:bounds}
0<\mu<2/\lambda_{max}
\end{equation}
Of course the simple LMS algorithm has the same guarantees for all estimated signals, sparse and dense alike. Using the sparsity assumption can increase the speed of convergence and yield much lower steady state estimation errors than the classical algorithms.
\subsection{Zero Attracting LMS}
The Zero Attracting LMS algorithm (ZA-LMS) \cite{Chen09} is a modification of the standard LMS algorithm that specializes in sparse system identification.  This algorithm follows the spirit of the equivalence of the $\ell_1$ and $\ell_0$ regularization problems in the batch case. Therefore the objective minimized at every iteration becomes
\begin{equation}
J_{ZA}(n) = \frac{1}{2} \abs{e(n)}^2 +\gamma\norm{\mathbf{w}(n)}_1
\end{equation}
for some parameter $\gamma$. Taking the gradient descent update one can adapt the LMS update scheme to the following
\begin{equation}
\mathbf{w}(n+1)=\mathbf{w}(n)+\mu e^*(n)\mathbf{x}(n)-\rho \mathrm{sgn}(\mathbf{w}(n))
\end{equation}
where $\rho=\mu\gamma$ and $\mathrm{sgn}(x)$ is the component wise sign function defined as 
\begin{equation}
\mathrm{sgn}(x)=
\begin{cases}
\frac{x}{\abs{x}} ,\quad x \neq 0 \\
0, \text{ otherwise}
\end{cases}
\end{equation}
It is clear that smaller coefficients of the estimated vector are quickly drawn to zero making the vector sparse while larger coefficients remain mostly unaffected for small values of $\rho$. Thus the update rule converges to sparse vectors.
\subsection{$\ell_0$-LMS}
$\ell_0$-LMS \cite{Gu2009} takes a different approach to sparse system identification by trying to minimize the objective
\begin{equation}
J_{\ell_0}(n) = \frac{1}{2} \abs{e(n)}^2 +\gamma\norm{\mathbf{w}(n)}_0
\end{equation}
Of course simply doing a gradient descent on the objective directly is not possible and in general the problem is known to be NP-hard. Instead the $\ell_0$ norm is approximated by 
\begin{equation}
\norm{\mathbf{w}(n)}_0 \approx \sum_{i=0}^{N-1} \left(1-e^{-\beta\abs{w_i(n)}}\right)
\end{equation}
The parameter $\beta$ here controls the quality of the approximation of the $\ell_0$ norm and as $\beta$ tends to infinity the formula becomes exact. Taking the gradient on the modified objective leads to the following update rule
\begin{equation}
\mathbf{w}(n+1)=\mathbf{w}(n)+\mu e^*(n)\mathbf{x}(n)-\rho \mathrm{sgn}(\mathbf{w}(n))e^{-\beta\abs{\mathbf{w}(n)}}
\end{equation}
where the exponentiation and the sign is applied element-wise. The same observations as in the previous algorithms apply here also. The difference is that the attraction to zero is even weaker for the coefficients that have large magnitudes so we expect that the convergence should be faster in general.
\section{New Online Algorithms}
\subsection{Selective Zero Attracting LMS} 
In the previous two sub sections we saw two regularized objectives of the standard LMS objective. In this paper we will try to solve a constrained version of the LMS objective. We will try to minimize 
\begin{equation}
J(n) = \frac{1}{2} \abs{e(n)}^2
\end{equation} 
but under the restriction that 
\begin{equation}
\norm{\mathbf{w}(n)}_0 \leq s
\end{equation}
where $s$, a positive integer less than $N$, is an upper bound on the sparsity of the vector under estimation that we know in advance. Let us define the operator $H_s$ that outputs a vector having zeros in all coefficients except for the ones with the $s$ largest absolute values that remain the same as in the input vector. For example if $\mathbf{x}_0=[2,-2,1,0]^T$ then $H_2(\mathbf{x}_0)=[2,-2,0,0]^T$. In case of ties we can take a conservative approach and allow all tying coefficients to be nonzero in the resulting vector so that $H_1(\mathbf{x}_0)=[2,-2,0,0]^T$. Thus $|\mathrm{support}(H_s(\mathbf{x}))| \geq s$ and therefore it is not guaranteed that the output will always be $s$-sparse. The operator will give as output vectors that are not $s$-sparse when there are multiple coefficients in the vector that their absolute value is equal to the $s$ largest absolute value in the vector. However, in most cases such ties will be nonexistent and the result will be an $s$-sparse vector. 
\par Given the definition of $H_s$ one could easily see a connection with $\ell_0$-LMS. Specifically we could relax the objective just like in the previous subsection. Here we will use however different $\beta_i$ for each coefficient. Let us approximate the $\ell_0$ norm as
\begin{equation}
\norm{\mathbf{w}(n)}_0 \approx \sum_{i=0}^{N-1} \left(1-e^{-\beta_i\abs{w_i(n)}}\right)
\end{equation}
Then if we want to make the estimate to converge to an $s$-sparse vector we can do the following: For the $s$ algebraically largest coefficients we will use $\beta_i=\infty$ whereas for all the others we will use $\beta_i=0$. This can be interpreted as the following penalty
\begin{equation}
{P_s(\mathbf{x})}_{i}=
\begin{cases}
0, \quad i \in \mathrm{support}(H_s(x))\\
\mathrm{sgn}(x_i),  \text{ otherwise} 
\end{cases}
\end{equation}
which then leads to the following update rule
\begin{equation} \label{eq:selective}
\mathbf{w}(n+1)=\mathbf{w}(n)+\mu e^*(n)\mathbf{x}(n)-\rho P_s(\mathbf{w}(n))
\end{equation}
This is the same concept of the $\ell_1$ penalization presented in \cite{Chen09} but applied only to the possibly superfluous coefficients given the a priori estimation of sparsity. We shall call this algorithm Selective Zero Attracting LMS.  Based on this fact we can prove a similar convergence result
\begin{theorem} \label{th:sza-lms}
Let us have a zero mean observation noise $v(n)$ independent of $\mathbf{x}(n)$ and given that $\mathbf{x}(n)$ and $\mathbf{w}(n)$ are independent. Let us also assume that $\mathbb{E}[\mathbf{x}(n)\mathbf{x}^H(n)]$ is constant over time, invertible and equal to $\mathbf{R}_x$. Then the algorithm described by (\ref{eq:selective}) converges in the mean sense provided that the condition of (\ref{eq:bounds}) holds. The limiting vector satisfies the equation
\begin{equation} \label{eq:bias}
\mathbb{E}[\mathbf{w}(\infty)]=\mathbf{w}-\frac{\rho}{\mu}\mathbf{R}_x^{-1}\mathbb{E}[P_s(\mathbf{w}(\infty))]
\end{equation}
\end{theorem}
The proof of the theorem can be found in Appendix \ref{ap:sza-lms} and is similar to the proof for the ZA-LMS. The interested reader can find an in depth analysis of a similar approximation scheme in \cite{al2016optimal}. The difference is in the choice of coefficients that get penalized and those who do not. In the update scheme presented we choose not to penalize the $s$ largest coefficients. In \cite{al2016optimal} the coefficients that do not get penalized are those who are greater than a predefined threshold.
\par As we can see in Equation (\ref{eq:bias}), the expected value of the the estimation does not converge necessarily to $\mathbf{w}$. In fact there might be a $O(\rho)$ deviation per coefficient just like in the simple Zero Attracting LMS. However, if $\mathbf{w}$ is an $s$ sparse vector and the algorithm identifies the support correctly then the bias for the leading $s$ coefficients should be eliminated as the penalty term will be zero for those coefficients, a property that the Zero Attracting LMS does not have. For the rest of the coefficients, unless the estimate for those does not converge exactly to 0 we will still incur the $O(\rho)$ deviation, which should be negligible for small values of $\rho$.
\subsection{Hard Threshold LMS} 
The contribution of this paper is the study of the properties of the following update scheme
\begin{equation} \label{eq:hard_lms}
\mathbf{w}(n+1)=H_s(\mathbf{w}(n)+\mu e^*(n)\mathbf{x}(n))
\end{equation}
\par It is easy to see the similarity of our proposed algorithm with the iterative hard thresholding algorithm studied in \cite{Blum09}, \cite{Blum10} and \cite{garg2009gradient}. There, since the algorithm is developed in a batch setting where all the data are known in advance, the relation between the observations $\mathbf{y}$ and the estimated vector $\mathbf{w}$ is $\mathbf{y}=\mathbf{A}\mathbf{w}$ where $\mathbf{A}$ is $M\times N$ matrix with $M<N$; thus the problem is undetermined. The update of the iterative hard thresholding under similar assumptions for the sparsity of $\mathbf{w}$ is
\begin{equation}
\mathbf{w}(n+1)=H_s(\mathbf{w}(n)+\mu \mathbf{A}^H\mathbf{e}(n))
\end{equation}
where $\mathbf{e}(n)=\mathbf{y}-\mathbf{A}\mathbf{w}(n)$. It must be noted that the complexity of implementing such an operator is still linear in $N$ as finding the $s$ largest value in a vector does not require sorting it first.
\par As a result it is clear that the proposed algorithm is closely related to the special case of iterative hard thresholding having $M=1$. It is also clear that we cannot use the rigorous proofs found in \cite{Blum09}, \cite{Blum10} and \cite{garg2009gradient} to show that the proposed algorithm also converges since for $M=1$ it is impossible to fulfill the strict properties needed. However, it is still possible to prove some interesting properties of the hard threshold operator. The main contribution of the operator is to let us focus our attention on the support of the estimated vector. If the algorithm does not provide a correct estimation of the support of the estimated vector then this could have a negative effect on the convergence of the algorithm. So one of the key properties that need to be studied is under which conditions is the estimation of the support using the hard threshold operator correct.
\begin{theorem} \label{th:strict}
Let $\mathbf{w}=[w_0,w_1,\dots,w_{N-1}]^T \in \mathbb{C}^N$ with $\norm{\mathbf{w}}_0=s$ and $\hat{\mathbf{w}}$ be an approximation. Let $q=\min_{w_i \neq 0} \abs{w_i}$. Then if $\norm{\mathbf{w}-\hat{\mathbf{w}}}_2^2 < \frac{q^2}{2}$ the following will be true 
\begin{equation} \label{eq:toprove}
\mathrm{support}(H_s(\hat{\mathbf{w}}))  = \mathrm{support}(\mathbf{w})
\end{equation}
\end{theorem}
The proof of the theorem is quite involved and can be found in Appendix \ref{ap:hard-lms}. The essence of the proof however is rather simple. In order to have the minimal error and still incorrectly specify the support of the vector, the error must be concentrated in two coefficients, one that belongs in $\mathrm{support}(\mathbf{w})$ and one that does not. The one coefficient that belongs to the correct support must end up having a smaller  magnitude than the one that should not. Since the first coefficient has at least magnitude $q$ in $\mathbf{w}$ and the other coefficient must have magnitude 0, the minimal error is achieved when both have magnitude $\frac{q}{2}$ in  $\hat{\mathbf{w}}$ which leads to the bound of the error that we have in the proof.
\par In order to understand the significance of the theorem we need to see some equivalent bounds having to do with the signal to error ratio that is needed so that the result in relation (\ref{eq:toprove}) still holds. The true vector $\mathbf{w}$ has $s$ nonzero values  each with an absolute value of at least $q$. Thus $\norm{\mathbf{w}}_2^2\geq sq^2$ and hence we need
\begin{equation} \label{eq:ser}
\mathrm{SER}=\frac{\norm{\mathbf{w}}_2^2}{\norm{\mathbf{w}-\hat{\mathbf{w}}}_2^2}> \frac{sq^2}{\frac{q^2}{2}}=2s
\end{equation}
\par Inequality (\ref{eq:ser}) is a necessary condition so that the required conditions of the theorem are true. Even if it is not sufficient it gives us the intuition that for small values of $s$ it will be easier to come up with an estimate $\hat{\mathbf{w}}$ for which relation (\ref{eq:toprove}) is true. On the other hand the conditions of Theorem~\ref{th:strict} are just sufficient for the relation (\ref{eq:toprove}) so in practice relation (\ref{eq:toprove}) could be true even with much lower signal to error ratios.
\par To further relax the conditions of our theorem we could allow the estimate to be less sparse. In order to do this we could use $H_d$ instead of $H_s$ with $N>d>s>0$ where $N$ is the size of the estimated vector. What happens here is a trade off. On the one hand, the result now is less attractive since we have more nonzero coefficients than what is actually needed and that may lead to excessive estimation error that could possibly be avoided. On the other hand, the estimation error of the input to the threshold operator can be greater without risking of loosing an element of $\mathrm{support}(\mathbf{w})$ after the application of the operator. The next theorem quantifies the gain in allowable estimation error.
\begin{theorem} \label{th:relaxed}
Let $\mathbf{w}$ be a vector in $\mathbb{C}^N$ with $\norm{\mathbf{w}}_0=s$ and $\hat{\mathbf{w}}$ be an approximation. Let $q=\min_{w_i \neq 0} \abs{w_i}$ and $d=s+\tau$ with $d<N$ and $\tau>0$ where $s$, $\tau$, $d$ are integers. Then if $\norm{\mathbf{w}-\hat{\mathbf{w}}}_2^2\leq q^2(1-\frac{1}{\tau+2})$ and $\norm{\hat{\mathbf{w}}}_0 \geq d $, the following will be true
\begin{equation} \label{eq:toprove_2}
\mathrm{support}(H_d(\hat{\mathbf{w}}))  \supseteq \mathrm{support}(\mathbf{w})
\end{equation} 
\end{theorem}
\par The proof of this theorem, found in the Appendix \ref{ap:relaxed}, is similar to the previous one. The difference in the result comes from the fact that $\tau+1$ coefficients that are not part of $\mathrm{support}(\mathbf{w})$ must have significant magnitudes in $\hat{\mathbf{w}}$ in order to miss a coefficient of $\mathrm{support}(\mathbf{w})$.  The analogous inequality of relation (\ref{eq:ser}) for this theorem is
\begin{equation}
\mathrm{SER} \geq \frac{s}{(1-\frac{1}{\tau+2})}
\end{equation}
which is less strict as we have expected.
\par Given the last theorem one can overcome the need to have an initialization that is too close to the vector to be estimated. If we have an estimate that has an error $\norm{\mathbf{w}-\hat{\mathbf{w}}}_2^2$ at most $q^2$, we can use the hard threshold operator to reduce its sparsity up to a degree that depends on the error without loosing an important coefficient and thus reducing the error in the process. Of course this is a worst case analysis and the conditions are sufficient but not necessary. Therefore in practice we should be able to to use the update rule of (\ref{eq:hard_lms}) without waiting to converge so close to the solution.
\subsection{Estimating Sparsity}
In some applications knowing an upper bound on sparsity, the parameter $s$ in our algorithms, may be an acceptable assumption. For example, in echo cancellation one can assume that there will be a small batch of tens of coefficients that are non-zero. In spectrum estimation we can calibrate $s$ based on prior knowledge about how many primary and secondary users of the spectrum are usually utilizing the spectrum. In general however, we would like our algorithm to adapt in different settings and therefore we need to be able to estimate the parameter $s$ in an online fashion.
\par  To achieve that we will assume that we have a knowledge of lower bound on $q$, the minimum magnitude of the non-zero coefficient in the estimated vector. One such lower bound could be the minimum magnitude required to consider the corresponding frequency occupied in the cognitive radio application. Let us call this value $q^*$. One naive way to estimate the sparsity could be to count the number of coefficients in the current estimate $\mathbf{w}(n)$ that have magnitude greater than $q^*$ and use this as an estimate for the sparsity. 
\par Unfortunately, the current estimation may not be suitable to use for sparsity estimation when the underlying spectrum is changing. For example, let us assume that the number of non zero coefficients increases. To increase our estimation of $s$ based on $\mathbf{w}(n)$ at least one coefficient's magnitude would need to go from 0 to above $q^*$ in a single iteration. Waiting for multiple iterations does not help if hard thresholding is used to remove negligible coefficients. But for such a significant jump to happen in a single iteration one would need either a small $q^*$ or a large $\mu$ both of which are undesirable as the first one reduces the accuracy of our sparsity estimate and the second one may make the estimation unstable. 
\par Instead we will try to approximate the error of our current estimate in order to construct a more suitable vector for the aforementioned procedure. The intuition behind this is that if we track an the error of our current estimate we can then use it to trigger increases in the parameter $s$ when the error increases significantly. Let $\mathbf{w}$ be once gain the true vector and $\mathbf{w}(n)$ our current estimate. We want to approximate $\mathbf{\bar{w}}(n)=\mathbf{w}(n)-\mathbf{w}$. From equation (\ref{eq:error}) we can get by taking the expectation and assuming that the noise has zero mean that
\begin{equation}
\mathbb{E}[e^*(n)\mathbf{x}(n)]=-\mathbb{E}[\mathbf{x}(n)\mathbf{x}^H(n)]\mathbb{E}[\bar{\mathbf{w}}(n)]
\end{equation}
$\mathbf{x}(n)$ correspond to rows of $\mathbf{\Phi}^*$. Since they are chosen uniformly at random we know that $\mathbb{E}[\mathbf{x}(n)\mathbf{x}^H(n)]=\mathbf{\Phi}^*\mathbf{\Phi}^T=\mathbf{I}_N$ where $\mathbf{I}_N$ is the $N\times N$ identity matrix. This equality is based on the properties of the IDFT matrix. Therefore the equation becomes
\begin{equation}
\mathbb{E}[e^*(n)\mathbf{x}(n)]=-\mathbb{E}[\bar{\mathbf{w}}(n)]
\end{equation}
Let $\mathbf{err}(n)$ be our approximation of $\bar{\mathbf{w}}(n)$. Ideally, we could take a set of new or even past measurements and calculate $e$ for them in every iteration to approximate the right hand side. This however would be wasteful and it would invalidate the online nature of the proposed algorithms. To avoid that we can reuse the estimations of the errors of the previous iterations. However, as our algorithm progresses, errors that were calculated many iterations ago in the past are not representative for our current estimate so they should be down-weighted compared to errors in recent iterations.  To overcome this we can take an exponentially weighted window average of the errors. Let $\lambda \in (0,1]$ be the forgetting factor of the window and $\mathbf{b}(n)=e^*(n)\mathbf{x}(n)$. Then we can write the following equations
\begin{align}
\kappa_{n+1} &= \lambda \kappa_n +1 \nonumber \\
\mathbf{err}(n+1) &= \left( 1 -\frac{1}{\kappa_{n+1}} \right) \mathbf{err}(n) -\frac{1}{\kappa_{n+1}}\mathbf{b}(n)
\end{align}
where $\mathbf{err}(0)$ is all zeros and  $\kappa_{0}$ is zero as well. 
\par In the end we will get a $\mathbf{w}'(n)=\mathbf{w}(n)-\mathbf{err}(n)$ and we will compare each coefficient magnitude and compare it to the threshold $q^*$. The number of coefficients that pass this test is the estimate of $s$. Optionally we can use the coefficients that pass the test as the support of $\mathbf{w}(n+1)$ in the next iteration in order to reduce the computational effort. 
\par The advantage of using this process instead of operating directly on  $\mathbf{w}(n)$ is that we can increase the importance of errors only for sparsity estimation and thus we avoid making our estimate unstable. In general we can even scale the importance of the error correction 
\begin{equation} \label{eq:xi}
\mathbf{w}'(n)=\mathbf{w}(n)-\xi\mathbf{err}(n)
\end{equation}
where $\xi$ is a positive real number. Holding $q^*$ fixed we can increase $\xi$ to make our sparsity estimation more sensitive to the error estimate. 
\section{Experimentation}
\begin{figure}[!t]
\centering
\includegraphics[scale=0.4]{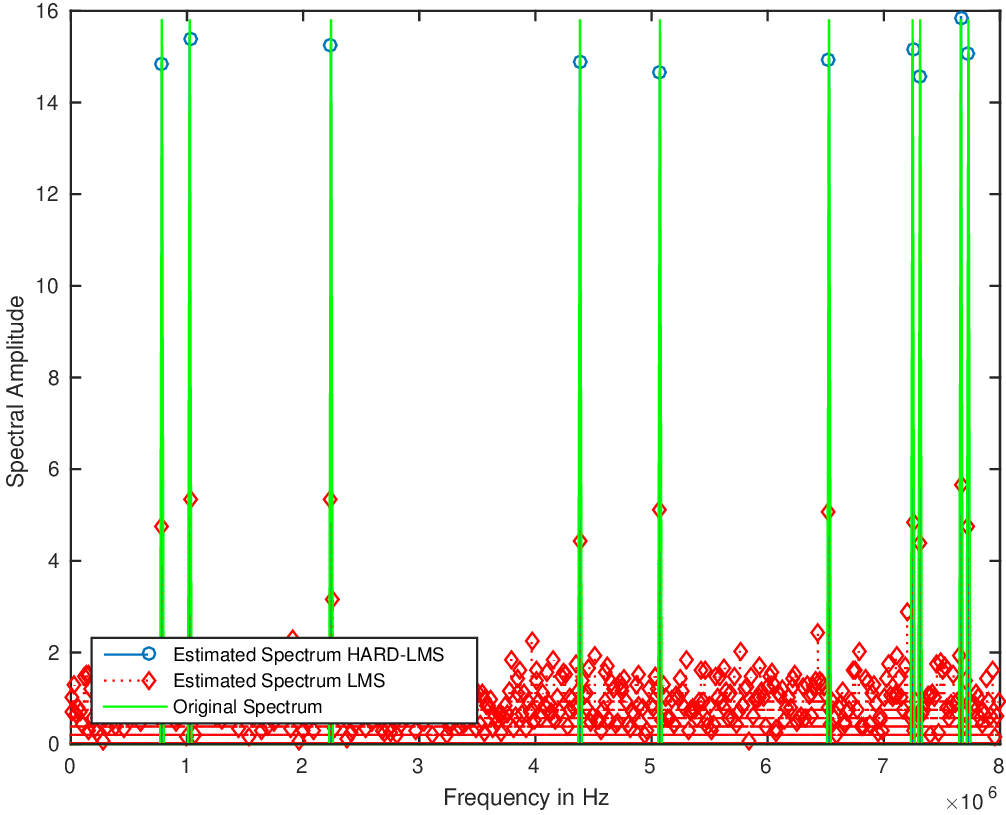}
\caption{Estimation of the spectrum with 20 non-zero coefficients by LMS and Hard Threshold LMS for $s=20$.}
\label{fig:radio}
\end{figure} 
In this section we will compare the performance of the various algorithms discussed previously. Let us first define the signals on which we will compare these algorithms. The signals of interest are going to be sums of sines affected by additive white Gaussian noise. Specifically the signals of interest here will have the form
\begin{equation}
g(n)= \sum_{i=1}^k A_i\sin(2\pi f_i t(n)) + v(n)
\end{equation}
where $k$ will be the number of the signals added, $f_i$ is the randomly chosen frequency of each sine, $A_i$ is the amplitude of each sine wave and $v(n)$ is the white zero mean noise. Therefore the spectrum of these signals will be sparse with $s=2k$ non-zero coefficients. The sampling positions $t(n)$ are spread uniformly in a time interval $T$ and the corresponding sampling frequency is equal to the Nyquist frequency. This results in a vector of $N$ samples per time interval $T$ out of which we will sample $M$ of those. Here we will assume for simplicity that $A_i=1$. 
\par The first thing we would like to show is that using sparsity aware techniques for spectrum estimation is a necessity when we are undersampling. We will therefore compare the spectrum estimations of the Hard Threshold LMS and the classical LMS. We will use the sum of $k=10$ sine waves of length $N=1000$ samples out of which we collect only $M=300$ samples corrupted by noise so that the SNR is equal to 20db. In order for the algorithms to converge we will make 10 iterations over the data. For the Hard Threshold LMS (HARD-LMS) algorithm we will use $s=20$ and we will refrain from thresholding during the first pass over the data. For both HARD-LMS and LMS we will use $\mu=1$. The results can be seen in Figure~\ref{fig:radio}. As we can clearly see the LMS algorithm does not converge to the true sparse vector that generated the measurements but simply to one of the many non-sparse solutions. In contrast HARD-LMS identified the support of the spectrum perfectly and the error is minimal compared to the one of LMS.
\begin{figure}[!t]
\centering
\includegraphics[scale=0.4]{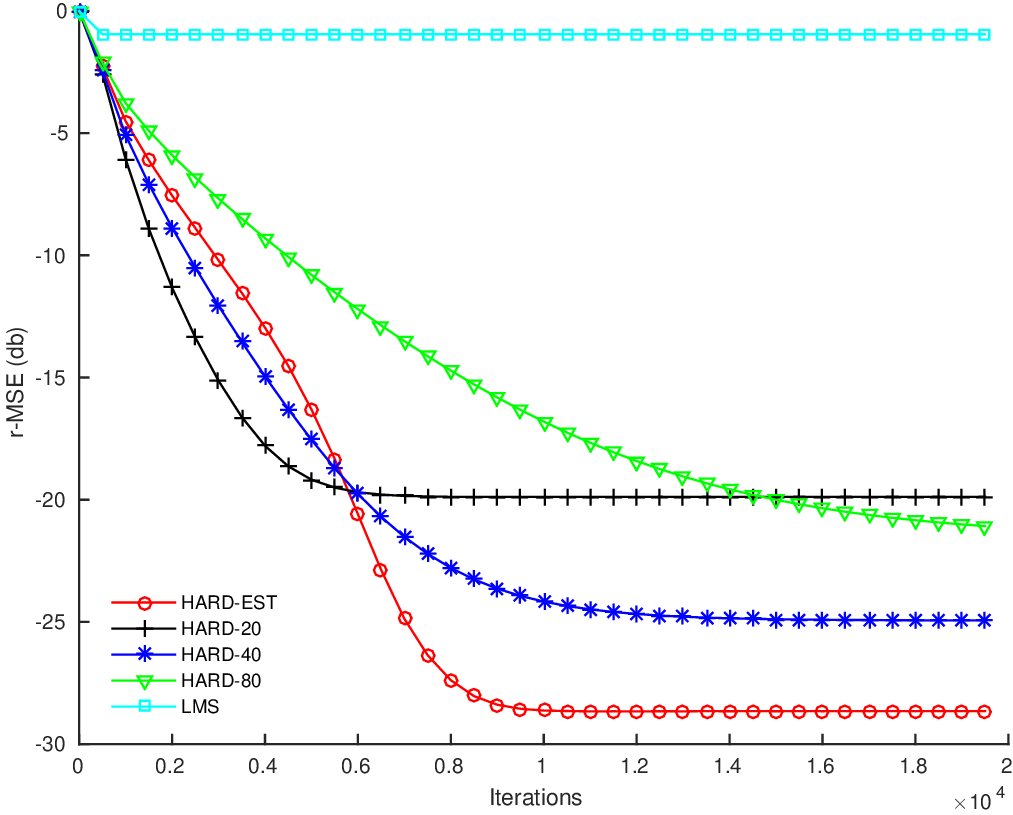}
\caption{Relative Mean Square Error of the spectrum estimation with 20 non-zero coefficients by various hard thresholding algorithms.}
\label{fig:hard_comparison}
\end{figure}  
\par Moreover, we would like to show experimentally how the parameter $s$ in our Hard Threshold LMS algorithm influences the speed of convergence as well as the steady state error. We set $N$ to be equal to 1000 and $M=200$ leading to a one to 5 undersampling ratio. We set $k$ to be equal to 10 and set the noise power so that the SNR of the observed samples is equal to 20db. We collect the $M$ samples and repeat them $K=100$ times to test the convergence of the algorithms. We repeat the whole process of choosing the different frequencies $f_i$ and the random $M$ samples for $R=200$ times. The algorithms that are compared are the following: The Hard Threshold LMS algorithm for values $s$ equal to 20 (HARD-20), 40 (HARD-40) and 80 (HARD-80) as well as the version of the Hard Threshold LMS with sparsity estimation (HARD-EST). For the sparsity estimation we use $\lambda=0.99$, $q^*$ equal to one tenth of the magnitude of any of the non zero coefficients of the spectrum (all have equal magnitude in this case) and $\xi=1$. For all those algorithms we refrain from hard thresholding for the first $2M$ samples so that we get a good enough approximation. Additionally, for all algorithms $\mu$ is set equal to 1. We also include the standard LMS (LMS) as a baseline comparison.
\par The results we get from Figure~\ref{fig:hard_comparison} are quite interesting. What is evaluated is the relative Mean Square Error (r-MSE). For every run of the algorithm and for every iteration we calculate
\begin{equation}
\mathrm{r-MSE}=\frac{\norm{\mathbf{w}-\mathbf{w}(n)}^2}{\norm{\mathbf{w}}^2}
\end{equation}
and then we take the average performance in db. As we can see selecting $s$ being exactly equal to the true sparsity is not always optimal. The algorithm for $s=20$ quickly converges to a suboptimal solution with high steady state error. This is because the algorithm has made a wrong estimation of the spectrum's support. In contrast allowing more non-zero coefficients allows the algorithm to include the true support of the spectrum as well as some superfluous coefficients. This allows both $s=40$ and $s=80$ to achieve much lower steady state errors. However, increasing the parameter $s$ will tend to significantly decrease the speed of convergence of the algorithm. On the other hand the hard thresholding algorithm with sparsity estimation by making better estimates of the true spectrum and using a conservative magnitude threshold gradually decreases the sparsity estimate in order to smoothly and quickly converge. This aligns with what we proved in the previous section especially with Theorem~\ref{th:relaxed}. Of course the classical LMS algorithm had no hope of finding the true spectrum as the problem is undetermined and LMS gets stuck in a non sparse spectrum that could give the observed measurements. Since HARD-EST achieved the best performance compared to all other methods we will compare it with other online sparsity aware algorithms from the literature. 
\begin{figure}[t]
\centering
\includegraphics[scale=0.4]{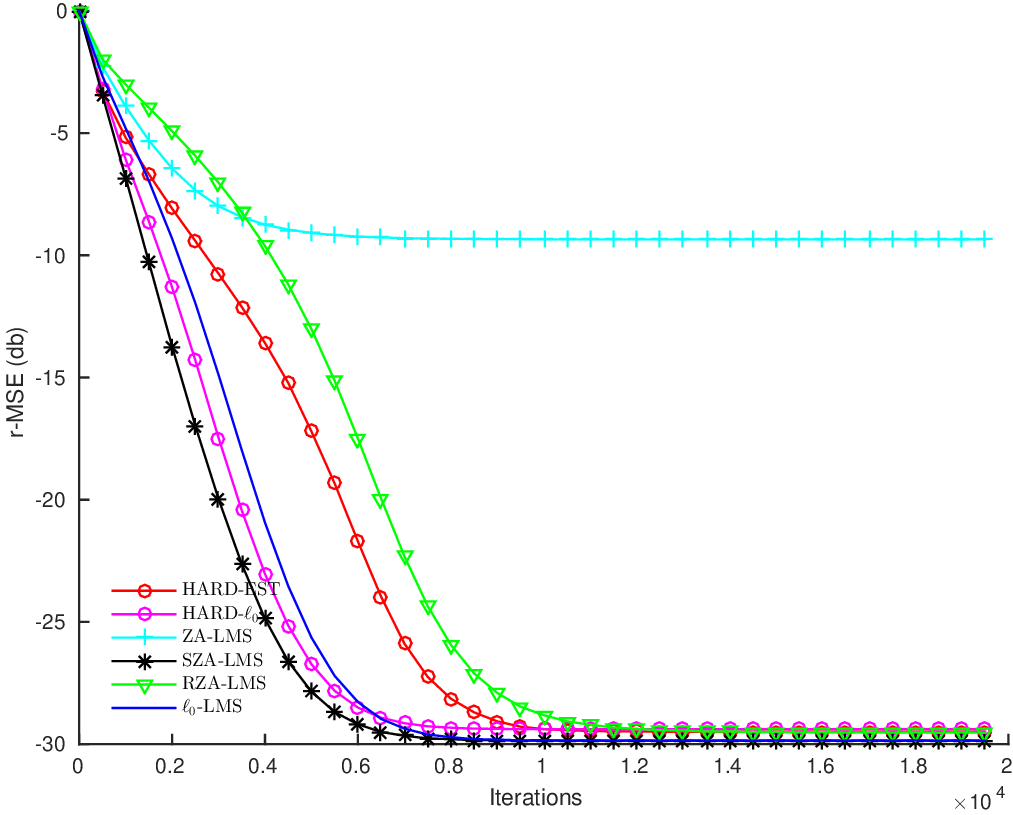}
\caption{Relative Mean Square Error of the spectrum estimation with 20 non-zero coefficients by sparsity aware algorithms proposed in the literature.}
\label{fig:sparse_comparison}
\end{figure}     
\par Specifically we will also compare with Zero Attracting LMS (ZA-LMS) and Reweighted Zero Attracting LMS (RZA-LMS) from \cite{Chen09} as well as with $\ell_0$-LMS from \cite{Gu2009}. We will set the parameter $\rho$ of ZA-LMS and RZA-LMS equal to $0.005$ and $\epsilon=2.25$. For the $\ell_0$-LMS we will set  $\beta=0.5$ and $\kappa\beta=0.005$. We will also include the Selective Zero Attracting LMS (SZA-LMS) that we proposed in this paper using the true sparsity $s=20$ and $\rho=0.005$. Additionally, the proposed hard thesholding scheme with sparsity estimation can be combined with other more complicated update rules to further improve performance. So for this experiment we will combine it with the update rule of $\ell_0$-LMS using the same parameters to show that we can get improved performance over each method alone. We shall call this algorithm HARD-$\ell_0$ and its update rule will be
\begin{equation}
\mathbf{w}(n+1)=H_s(\mathbf{w}(n)+\mu e^*(n)\mathbf{x}(n)-\rho \mathrm{sgn}(\mathbf{w}(n))e^{-\beta\abs{\mathbf{w}(n)}})
\end{equation} 
where $s$ is estimated the same way as for HARD-EST. For HARD-EST we will refrain from hard thesholding for the first $M$ samples and for HARD-$\ell_0$ for the first $2M$ samples. The experimental settings are the same as in the previous experiment.

\par The results can be seen in Figure~\ref{fig:sparse_comparison} where we show again the r-MSE. Clearly, we can see that all algorithms manage to reach nearly the same level of r-MSE after some iterations with SZA-LMS and $\ell_0$-LMS achieving a slightly smaller r-MSE than the other methods. So it makes sense to compare them in terms of speed of convergence. The fastest convergence belongs to SZA-LMS. SZA-LMS has ground truth knowledge of the sparsity of the vector just like the hard thresholding algorithms in the previous experiments but uses it slowly but steadily in order not to reduce coefficients of the true support to 0.  Then we have HARD-$\ell_0$ which combines the quickly convergent update rule of $\ell_0$-LMS with hard thresholding improving the convergence speed of an already fast algorithm like $\ell_0$-LMS.  Then $\ell_0$-LMS with an update rule that uses two parameters to tune the amount of zero attraction to use for each coefficient manages to converge faster than the simpler HARD-EST algorithm. HARD-EST then manages to converge faster than RZA-LMS. Finally the simple ZA-LMS algorithm fails to achieve a low steady state error. 
\par
\begin{figure}[t]
\centering
\includegraphics[scale=0.4]{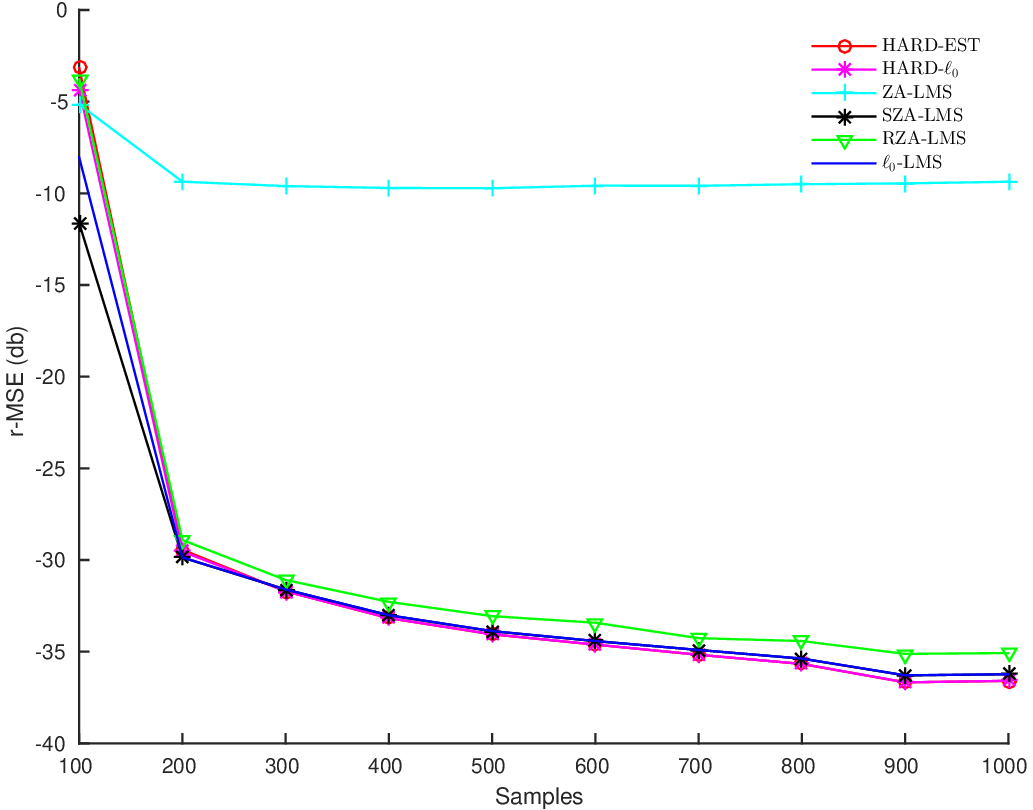}
\caption{Relative Mean Square Error of the spectrum estimation with 20 non-zero coefficients with varying amount of observations.}
\label{fig:number_of_samples}
\end{figure}

 The third experiment that we will present has to with the robustness of the proposed procedures with varying degrees of undersampling. We evaluate the sparsity aware algorithms for $M=100, 200, \dots 1000$ samples where 1000 samples corresponds to the full measurement of the signal. In each setting we take 50 instantiations of the random sampling procedure. Then we calculate the steady state r-MSE after $K=50$ iterations over all the available measurements. The results are shown in Figure~\ref{fig:number_of_samples}. We compare once again the same algorithms with the same parameters as in the previous experiment. We can clearly see that reducing the number of samples to 100 which corresponds to a 1 over 10 undersampling ratio is prohibitive for all algorithms except maybe SZA-LMS which has ground truth knowledge. However, once we get to 200 samples, which was the case in the previous experiment all algorithms improve their predictions considerably. Adding even more samples leads to better performance although with diminishing returns. One pattern that may seem counter-intuitive is that the hard thresholding algorithms, HARD-EST and HARD-$\ell_0$, manage to outperform all other methods by a small margin after 200 samples which is in contrast to what we saw in the previous experiment. The reasoning behind this is that HARD-EST and HARD-$\ell_0$ have no misalignment with the ground truth for the coefficients that are exactly zero. In contrast for the other methods these weights oscillate around zero  due to the zero attraction term for the same reasons we analyzed for the case of SZA based on Equation (\ref{eq:bias}). This difference in performance is quite small so it is only observable when HARD-EST and HARD-$\ell_0$ have reached their optimal performance.

\par In the fourth experiment we are going to validate that HARD-EST is capable of tracking changing sparsity patterns in real time. In this experiment, the incoming signal will change over time. At first, the incoming signal consists of 10 sine waves just like before. The pattern of $N=1000$ samples is repeated 150 times. Then the incoming signal is augmented with another 10 sine waves of different frequency. Then the new pattern is repeated for 150 times. The incoming signal is split in non overlapping windows of $N$ samples and we randomly sample $M=200$ measurements corrupted by additive noise in each window. The SNR is 20db.  
\par
\begin{figure}[!t]
\centering
\includegraphics[scale=0.4]{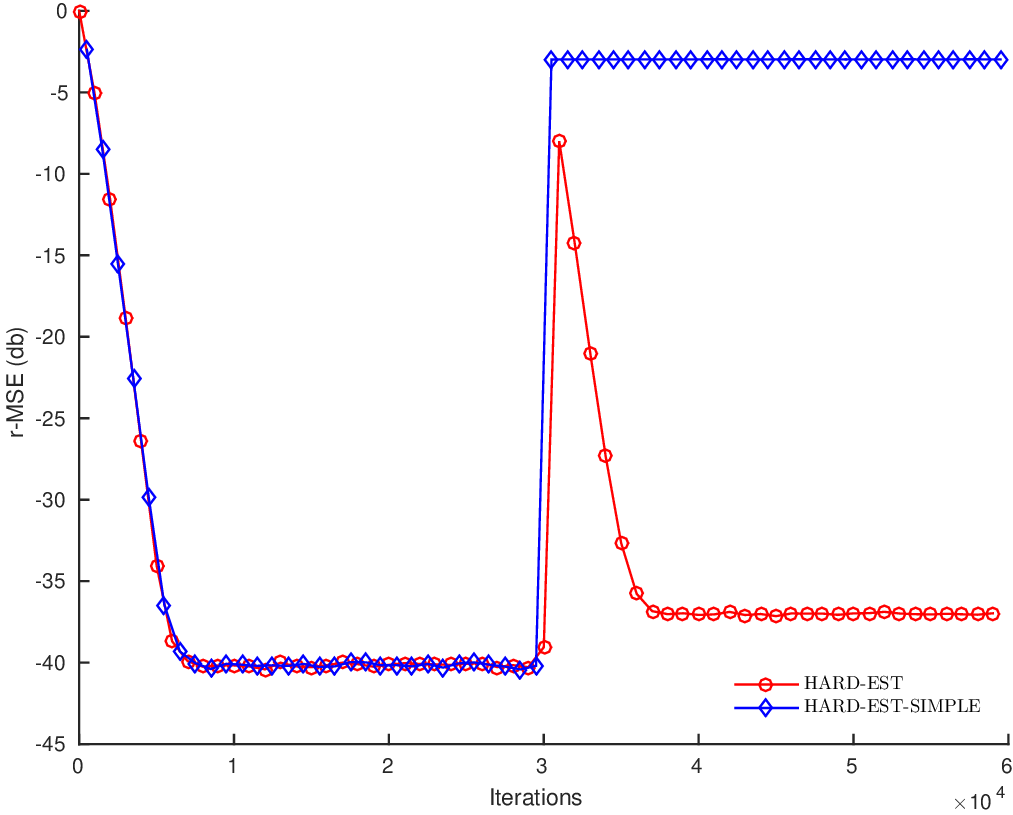}
\caption{Adaptation of the HARD-EST and HARD-EST-SIMPLE when the incoming signal changes from 10 sine waves to 20 sine waves after 30000 samples.}
\label{fig:hard_est_adaptation}
\end{figure}

 To help HARD-EST perform well in scenarios where the incoming signal is changing abruptly we are going to change the algorithms configuration. We are going to set $\lambda=0.98$, $q^*$ equal to one hundredth of the magnitude of any of the non zero coefficients of the spectrum (again all the coefficients have equal magnitude) and $\xi=20$.  The smaller $\lambda$ allows us to forget previous error estimates more quickly whereas the combination of the smaller $q^*$ and the higher $\xi$ allows us to adapt more quickly to changes in the sparsity of the spectrum. We can see the results in Figure~\ref{fig:hard_est_adaptation}. The algorithm converges very close to the true spectrum using the new samples it gets in every window. When the change in the spectrum happens the estimation is initially far away from the new spectrum. Then, the increased error estimates trigger the increase of the estimated sparsity from 20 towards 40 non zero coefficients. This allows the estimate to adapt to the less sparse spectrum and eventually converge to it. The r-MSE in the end is higher than before the spectrum change but this is to expected since the spectrum now is less sparse.
\par To understand the effect of the parameter $\xi$ in Equation~\ref{eq:xi} and the significance of our sparsity estimation procedure  we add an additional algorithm HARD-EST-SIMPLE which is the same as HARD-EST with the only difference being that for HARD-EST-SIMPLE we set $\xi=0$. In this setting the sparsity estimation is successful in the first half of the simulation yielding the same approximation error with HARD-EST. However, in the second half while HARD-EST manages to increase its estimate $s$ to 40, HARD-EST-SIMPLE does not manage to adapt resulting in $s$ being equal to 20 in the second half as well and in an r-MSE of -3 db. Therefore, it is clear that, when the underlying patterns of sparsity are changing, setting a positive value for $\xi$ is crucial for the adaptation of the sparsity estimation.
\section{Conclusion}
In this paper we studied the problem of online sparse spectrum estimation for cognitive radio applications using sub-Nyquist sampling rates. To solve the problem, we analyzed the properties of two algorithms that try to minimize the squared error at each iteration while maintaining the $\ell_0$ norm of the estimated vector under a predefined threshold $s$. Moreover, we analyzed the convergence properties of the Selective Zero Attracting LMS as well as the properties of the Hard Thresholding operator. Specifically, we proved that if our current estimation is close enough to the solution we can use the Hard Thresholding operator to reduce the error without risking to loose an important coefficient of the spectrum especially when we allow the operator to use more non-zero coefficients. Additionally, we proposed a way to estimate in an adaptive way the parameter $s$ so that the estimation can gradually become sparser without misspecifying the support of the estimated spectrum. Further, in the experimentation section we analyzed the importance of the parameter $s$ for the steady state error as well as the speed of convergence. Then we compared our results with other online sparsity aware algorithms in the literature. We also showed that the two proposed algorithms have robust performance even when the sampling rate is low and that they can produce even better estimates when the number of samples increases. Finally, we showed experimentally that the proposed sparsity estimation technique is robust to signal changes.
\par Of course spectrum estimation for cognitive radio applications is only one of the many possible applications of the proposed algorithms. Obviously an a priori estimation of the sparsity of the estimated vector may not be available in all use cases, even though we showed that this estimate must not be exact in order to actually take benefit. However, there are other use cases where the algorithms proposed here could make a difference. The standard LMS algorithm has been used in many adaptive machine learning tasks like neural network training and others as discussed in \cite{Theod15} so taking advantage of sparsity could be advantageous. For example, in the case of training a perceptron with an abundance of available features one could begin training with all the features but then proceed to use one of the proposed algorithms to impose feature selection through sparsity. By increasing the imposed sparsity one can then train several classifiers and then compare them using criteria like the Bayesian information criterion.

\appendices
\section{Proof of Theorem~\ref{th:sza-lms}} \label{ap:sza-lms}
\begin{IEEEproof}
Let us define $\bar{\mathbf{w}}(n)$ as the difference between the estimation $\mathbf{w}(n)$ and the true vector $\mathbf{w}$. Subtracting $\mathbf{w}$ from both sides of the equation (\ref{eq:selective}) gives
\begin{equation} \label{eq:update}
\begin{aligned}
\bar{\mathbf{w}}(n+1)&=\mathbf{w}(n)+\mu e^*(n)\mathbf{x}(n)-\mathbf{w} -\rho P_s(\mathbf{w}(n))+v(n)\\
&=\bar{\mathbf{w}}(n)+\mu e^*(n) \mathbf{x}(n) -\rho P_s(\mathbf{w}(n))+v(n)\\
\end{aligned} 
\end{equation}
\par After some calculations, which are the same as in the case of the classical LMS, we have that
\begin{equation}
e^*(n)\mathbf{x}(n)=-\mathbf{x}(n)\mathbf{x}^H(n)\bar{\mathbf{w}}(n)+v(n)\mathbf{x}(n)
\end{equation}
Taking the mean under the independence assumptions made and given that the observation noise mean is zero will yield
\begin{equation}
\mathbb{E}[e^*(n)\mathbf{x}(n)]=-\mathbf{R}_x\mathbb{E}[\bar{\mathbf{w}}(n)]
\end{equation}
Then from equation (\ref{eq:update}) we obtain 
\begin{equation}
\mathbb{E}[\bar{\mathbf{w}}(n+1)]=(\mathbf{I}_N-\mu \mathbf{R}_x)\mathbb{E}[\bar{\mathbf{w}}(n)]-\rho \mathbb{E}[P_s(\mathbf{w}(n))]
\end{equation}
where $\mathbf{I}_N$ is the $N\times N$ identity matrix. Given the bound in (\ref{eq:bounds}) the algebraically largest eigenvalue of $\mathbf{I}_N-\mu\mathbf{R}_x$ is less than one. Further the term induced by the penalty is bounded by the vectors $-\rho\mathbf{1}$ and $\rho\mathbf{1}$ where $\mathbf{1}$ is the vector of $\mathbb{R}^N$ whose every element is one. Thus we can conclude that the  $\mathbb{E}[\bar{\mathbf{w}}(n)]$ converges and as a result so does  $\mathbb{E}[\mathbf{w}(n)]$. Therefore the algorithm provided by equation (\ref{eq:selective}) converges. The limiting vector cannot be found in a closed form but is guaranteed to be the solution of equation (\ref{eq:bias}).
\end{IEEEproof}
\section{Proof of Theorem~\ref{th:strict}} \label{ap:hard-lms}
\begin{IEEEproof}
The proof will be completed in three distinct cases.
\par (i) First, we assume that $\norm{H_s(\hat{\mathbf{w}})}_0<s$ which can be true only if $\norm{\hat{\mathbf{w}}}_0<s$. We can easily see that, since  $\norm{\mathbf{w}}_0=s$, there is at least one coefficient index $i$ such that $\hat{w}_i=0$ and $w_i\neq 0$, which from the hypothesis also means that $\abs{w_i}\geq q$. As a result we have that $$\norm{\mathbf{w}-\hat{\mathbf{w}}}_2^2 \geq \abs{w_i-\hat{w}_i}^2= \abs{w_i}^2 \geq q^2$$ which contradicts the hypothesis; so this case is impossible.
\par (ii) Now we have that $\norm{H_s(\hat{\mathbf{w}})}_0 = s$. Let us assume that relation (\ref{eq:toprove}) does not hold. Then since the two sets have the same number of nonzero elements, it is clear that there is a coefficient index $\ell\in \mathrm{support}(\mathbf{w})$ but $\ell \notin \mathrm{support}(H_s(\hat{\mathbf{w}}))$ and a coefficient index $k$ so that $k \in \mathrm{support}(H_s(\hat{\mathbf{w}}))$ but $k \notin \mathrm{support}(\mathbf{w})$. We directly know that $w_k=0$ and that $\abs{w_\ell}\geq q$. We can also deduce that $\abs{\hat{w}_k} > \abs{\hat{w}_\ell}$ since $k$ belongs in $\mathrm{support}(H_s(\hat{\mathbf{w}}))$ but $\ell$ does not. Then, for the error norm we have 
\begin{equation*}
\norm{\mathbf{w}-\hat{\mathbf{w}}}_2^2 \geq \abs{w_k-\hat{w}_k}^2+\abs{w_\ell-\hat{w}_\ell}^2
\end{equation*}
Since $\abs{w_k-\hat{w}_k}^2 = \abs{\hat{w}_k}^2 > \abs{\hat{w}_\ell}^2 $, it follows that 
\begin{equation*}
\norm{\mathbf{w}-\hat{\mathbf{w}}}_2^2 > 2\abs{\hat{w}_\ell}^2-w^*_\ell \hat{w}_\ell -w_\ell \hat{w}^*_\ell + \abs{w_\ell}^2
\end{equation*} 
Therefore we can also write that
\begin{equation*}
\norm{\mathbf{w}-\hat{\mathbf{w}}}_2^2 >\min_{\hat{w}_\ell \in \mathbb{C}}{2\abs{\hat{w}_\ell}^2-w^*_\ell \hat{w}_\ell -w_\ell \hat{w}^*_\ell + \abs{w_\ell}^2} 
\end{equation*} 
The minimum value of the RHS is attained for $\hat{w}_\ell=\frac{w^*_\ell}{2}$ and equals $\frac{\abs{w_\ell}^2}{2}$; hence 
\begin{equation}
\norm{\mathbf{w}-\hat{\mathbf{w}}}_2^2 >\frac{\abs{w_\ell}^2}{2} \geq \frac{q^2}{2} 
\end{equation} 
This once again contradicts the hypothesis and so relation (\ref{eq:toprove}) is true in this case.
\par (iii) Finally, we assume that $\norm{H_s(\hat{\mathbf{w}})}_0 > s$. This can happen only if there are ties for the s largest absolute values in $\hat{\mathbf{w}}$. Let us denote as $B$ the set of tying coefficients, $A= \mathrm{support}(H_s(\hat{\mathbf{w}})) \setminus B$ and finally $C=(\mathrm{support}(H_s(\hat{\mathbf{w}}))^c$. It is evident that $\abs{A} \leq s-1$. We shall prove that this case is impossible. There are two subcases:
\par (a) $B \cap \mathrm{support}(\mathbf{w})= \emptyset$. Since $\abs{A} \leq s-1$ and $\norm{w}_0=s$, $\mathrm{support}(\mathbf{w})$ must have an element in common with $C$. Let us call that element $\ell$. Let us also take an element $k$ from $B$.  Then just like in the second case $\abs{\hat{w}_k} > \abs{\hat{w}_\ell}$ since $k$ belongs in $\mathrm{support}(H_s(\hat{\mathbf{w}}))$ but $\ell$ does not. Following the rest of the steps in case (ii) we reach a contradiction.
\par (b) $B \cap \mathrm{support}(\mathbf{w})\neq \emptyset$. Let $\ell$ a common element of the two sets. Since $\norm{H_s(\hat{\mathbf{w}})}_0 >\norm{\mathbf{w}}_0$ there is an element $k$ so that $k \in \mathrm{support}(H_s(\hat{\mathbf{w}}))$ but $k \notin \mathrm{support}(\mathbf{w})$. Since $\ell$ is one of the indexes tying for the last spot, we have $\abs{\hat{w}_k} \geq \abs{\hat{w}_\ell}$. Following the steps of case (ii) yields $\norm{\mathbf{w}-\hat{\mathbf{w}}}_2^2 \geq \frac{\abs{w_\ell}^2}{2} \geq \frac{q^2}{2} $ and therefore we get a contradiction.
\end{IEEEproof}
\section{Proof of the Theorem~\ref{th:relaxed}} \label{ap:relaxed}
\begin{IEEEproof}
Let us assume that relation (\ref{eq:toprove_2}) does not hold. Just like in the proof of Theorem~\ref{th:strict} it is clear that there is a coefficient index so that $\ell\in \mathrm{support}(\mathbf{w})$ but $\ell \notin \mathrm{support}(H_d(\hat{\mathbf{w}}))$. This time however the set $\mathrm{support}(H_d(\hat{\mathbf{w}}))$ has at least $d=s+\tau$ elements but $\mathrm{support}(\mathbf{w})$ has at most $s-1$ elements that could exist in  $\mathrm{support}(H_d(\hat{\mathbf{w}}))$. As a result we are sure that there are at least $\tau+1$ indexes $k_i$ so that $k_i \in \mathrm{support}(H_s(\hat{\mathbf{w}}))$ but $k_i \notin \mathrm{support}(\mathbf{w})$. Once again we know that $w_{k_i}=0$ and that $\abs{w_\ell}\geq q$ and we can deduce that $\abs{\hat{w}_{k_i}} > \abs{\hat{w}_\ell}$ since $k_i$ exists in $\mathrm{support}(H_d(\hat{\mathbf{w}}))$ but $\ell$ does not.
\par Like in the the proof of Theorem~\ref{th:strict} we can deduce about the error norm that $$\norm{\mathbf{w}-\hat{\mathbf{w}}}_2^2 \geq \sum_{i=1}^{\tau+1}{\abs{w_{k_i}-\hat{w}_{k_i}}^2}+\abs{w_\ell-\hat{w}_\ell}^2$$
We bound the first term just like in the previous proof so that it becomes $$\sum_{i=1}^{\tau+1}{\abs{w_{k_i}-\hat{w}_{k_i}}^2} = \sum_{i=1}^{\tau+1}{\abs{\hat{w}_{k_i}}^2} \geq (\tau+1) \abs{\hat{w}_\ell}^2$$ Thus, we end up $$\norm{\mathbf{w}-\hat{\mathbf{w}}}_2^2 > (\tau+2){\hat{w}_\ell}^2-2w_\ell \hat{w}_\ell + \abs{w_\ell}^2 $$
\par Taking the minimum on the right side with respect to $\hat{w}_\ell$ will lead once again to finding the minimum value of a quadratic function. The minimum is found for $\hat{w}_\ell= \frac{w_\ell}{\tau+2}$ and equals to ${w_\ell}^2(1-\frac{1}{\tau+2})$; hence
$$\norm{\mathbf{w}-\hat{\mathbf{w}}}_2^2 >{w_\ell}^2(1-\frac{1}{\tau+2}) \geq q^2(1-\frac{1}{\tau+2}) $$
which once again contradicts the hypothesis so the proof is completed.
\end{IEEEproof}

\section*{Acknowledgment}
We wish to thank the anonymous reviewers whose constructive comments helped us improve this paper.

\bibliographystyle{IEEEtran}
\bibliography{IEEEabrv,publication}
\end{document}